\documentclass[twocolumn,aps,prl,showpacs,preprintnumbers,superscriptaddress,floatfix]{revtex4}
\usepackage{rotating}% Include figure files
\usepackage{graphicx}% Include figure files
\usepackage{dcolumn}% Align table columns on decimal point
\usepackage{bm}% bold math

\begin{document}

%
% Useful Phobos Definitions
%

\newcommand{\snn}{\sqrt{s_{_{NN}}}}
\newcommand{\ee}{$e^{+}+e^{-}$}
\newcommand{\pbarp}{pp/\overline{p}p}
\newcommand{\zvtx}{z_{vtx}}
\newcommand{\npart}{N_{\rm part}}
\newcommand{\ncoll}{N_{\rm coll}}
\newcommand{\avenp}{\langle\npart\rangle}
\newcommand{\avenc}{\langle\ncoll\rangle}
\newcommand{\half}{\frac{1}{2}}
\newcommand{\halfnp}{(\half\avenp)}
\newcommand{\nch}{N_{\rm ch}}
\newcommand{\etazero}{\eta = 0}
\newcommand{\etaone}{|\eta| < 1}
\newcommand{\dndeta}{d\nch/d\eta}
\newcommand{\dndetazero}{\dndeta|_{\etazero}}
\newcommand{\dndetaone}{\dndeta|_{\etaone}}
\newcommand{\dndetanp}{\dndeta / \halfnp}
\newcommand{\dndetaonp}{\dndeta / \np}
\newcommand{\dndetazeronp}{\dndetazero / \halfnp}
\newcommand{\dndetaonenp}{\dndetaone / \halfnp}
\newcommand{\Pt}{p$_{_{\rm T}}$}

\preprint{Version 1.2}

\title{Scaling properties in bulk and \Pt-dependent particle production
near midrapidity \\ in relativistic heavy ion collisions.}

\author{B.Alver}
\affiliation{Laboratory for Nuclear Science, Massachusetts Institute of Technology, Cambridge, MA 02139-4307, USA}
\author{B.B.Back}
\affiliation{Physics Division, Argonne National Laboratory, Argonne, IL 60439-4843, USA}
\author{M.D.Baker}
\affiliation{Chemistry and C-A Departments, Brookhaven National Laboratory, Upton, NY 11973-5000, USA}
\author{M.Ballintijn}
\affiliation{Laboratory for Nuclear Science, Massachusetts Institute of Technology, Cambridge, MA 02139-4307, USA}
\author{D.S.Barton}
\affiliation{Chemistry and C-A Departments, Brookhaven National Laboratory, Upton, NY 11973-5000, USA}
\author{R.R.Betts}
\affiliation{Department of Physics, University of Illinois at Chicago, Chicago, IL 60607-7059, USA}
\author{R.Bindel}
\affiliation{Department of Chemistry, University of Maryland, College Park, MD 20742, USA}
\author{W.Busza}
\affiliation{Laboratory for Nuclear Science, Massachusetts Institute of Technology, Cambridge, MA 02139-4307, USA}
\author{Z.Chai}
\affiliation{Chemistry and C-A Departments, Brookhaven National Laboratory, Upton, NY 11973-5000, USA}
\author{V.Chetluru}
\affiliation{Department of Physics, University of Illinois at Chicago, Chicago, IL 60607-7059, USA}
\author{E.Garc\'{\i}a}
\affiliation{Department of Physics, University of Illinois at Chicago, Chicago, IL 60607-7059, USA}
\author{T.Gburek}
\affiliation{Institute of Nuclear Physics PAN, Krak\'{o}w, Poland}
\author{K.Gulbrandsen}
\affiliation{Laboratory for Nuclear Science, Massachusetts Institute of Technology, Cambridge, MA 02139-4307, USA}
\author{J.Hamblen}
\affiliation{Department of Physics and Astronomy, University of Rochester, Rochester, NY 14627, USA}
\author{I.Harnarine}
\affiliation{Department of Physics, University of Illinois at Chicago, Chicago, IL 60607-7059, USA}
\author{C.Henderson}
\affiliation{Laboratory for Nuclear Science, Massachusetts Institute of Technology, Cambridge, MA 02139-4307, USA}
\author{D.J.Hofman}
\affiliation{Department of Physics, University of Illinois at Chicago, Chicago, IL 60607-7059, USA}
\author{R.S.Hollis}
\affiliation{Department of Physics, University of Illinois at Chicago, Chicago, IL 60607-7059, USA}
\author{R.Ho\l y\'{n}ski}
\affiliation{Institute of Nuclear Physics PAN, Krak\'{o}w, Poland}
\author{B.Holzman}
\affiliation{Chemistry and C-A Departments, Brookhaven National Laboratory, Upton, NY 11973-5000, USA}
\author{A.Iordanova}
\affiliation{Department of Physics, University of Illinois at Chicago, Chicago, IL 60607-7059, USA}
\author{J.L.Kane}
\affiliation{Laboratory for Nuclear Science, Massachusetts Institute of Technology, Cambridge, MA 02139-4307, USA}
\author{P.Kulinich}
\affiliation{Laboratory for Nuclear Science, Massachusetts Institute of Technology, Cambridge, MA 02139-4307, USA}
\author{C.M.Kuo}
\affiliation{Department of Physics, National Central University, Chung-Li, Taiwan}
\author{W.Li}
\affiliation{Laboratory for Nuclear Science, Massachusetts Institute of Technology, Cambridge, MA 02139-4307, USA}
\author{W.T.Lin}
\affiliation{Department of Physics, National Central University, Chung-Li, Taiwan}
\author{C.Loizides}
\affiliation{Laboratory for Nuclear Science, Massachusetts Institute of Technology, Cambridge, MA 02139-4307, USA}
\author{S.Manly}
\affiliation{Department of Physics and Astronomy, University of Rochester, Rochester, NY 14627, USA}
\author{A.C.Mignerey}
\affiliation{Department of Chemistry, University of Maryland, College Park, MD 20742, USA}
\author{R.Nouicer}
\affiliation{Department of Physics, University of Illinois at Chicago, Chicago, IL 60607-7059, USA}
\author{A.Olszewski}
\affiliation{Institute of Nuclear Physics PAN, Krak\'{o}w, Poland}
\author{R.Pak}
\affiliation{Chemistry and C-A Departments, Brookhaven National Laboratory, Upton, NY 11973-5000, USA}
\author{C.Reed}
\affiliation{Laboratory for Nuclear Science, Massachusetts Institute of Technology, Cambridge, MA 02139-4307, USA}
\author{E.Richardson}
\affiliation{Department of Chemistry, University of Maryland, College Park, MD 20742, USA}
\author{C.Roland}
\affiliation{Laboratory for Nuclear Science, Massachusetts Institute of Technology, Cambridge, MA 02139-4307, USA}
\author{G.Roland}
\affiliation{Laboratory for Nuclear Science, Massachusetts Institute of Technology, Cambridge, MA 02139-4307, USA}
\author{J.Sagerer}
\affiliation{Department of Physics, University of Illinois at Chicago, Chicago, IL 60607-7059, USA}
\author{I.Sedykh}
\affiliation{Chemistry and C-A Departments, Brookhaven National Laboratory, Upton, NY 11973-5000, USA}
\author{C.E.Smith}
\affiliation{Department of Physics, University of Illinois at Chicago, Chicago, IL 60607-7059, USA}
\author{M.A.Stankiewicz}
\affiliation{Chemistry and C-A Departments, Brookhaven National Laboratory, Upton, NY 11973-5000, USA}
\author{P.Steinberg}
\affiliation{Chemistry and C-A Departments, Brookhaven National Laboratory, Upton, NY 11973-5000, USA}
\author{G.S.F.Stephans}
\affiliation{Laboratory for Nuclear Science, Massachusetts Institute of Technology, Cambridge, MA 02139-4307, USA}
\author{A.Sukhanov}
\affiliation{Chemistry and C-A Departments, Brookhaven National Laboratory, Upton, NY 11973-5000, USA}
\author{A.Szostak}
\affiliation{Chemistry and C-A Departments, Brookhaven National Laboratory, Upton, NY 11973-5000, USA}
\author{M.B.Tonjes}
\affiliation{Department of Chemistry, University of Maryland, College Park, MD 20742, USA}
\author{A.Trzupek}
\affiliation{Institute of Nuclear Physics PAN, Krak\'{o}w, Poland}
\author{G.J.van~Nieuwenhuizen}
\affiliation{Laboratory for Nuclear Science, Massachusetts Institute of Technology, Cambridge, MA 02139-4307, USA}
\author{S.S.Vaurynovich}
\affiliation{Laboratory for Nuclear Science, Massachusetts Institute of Technology, Cambridge, MA 02139-4307, USA}
\author{R.Verdier}
\affiliation{Laboratory for Nuclear Science, Massachusetts Institute of Technology, Cambridge, MA 02139-4307, USA}
\author{G.Veres}
\affiliation{Laboratory for Nuclear Science, Massachusetts Institute of Technology, Cambridge, MA 02139-4307, USA}
\author{P.Walters}
\affiliation{Department of Physics and Astronomy, University of Rochester, Rochester, NY 14627, USA}
\author{E.Wenger}
\affiliation{Laboratory for Nuclear Science, Massachusetts Institute of Technology, Cambridge, MA 02139-4307, USA}
\author{D.Willhelm}
\affiliation{Department of Chemistry, University of Maryland, College Park, MD 20742, USA}
\author{F.L.H.Wolfs}
\affiliation{Department of Physics and Astronomy, University of Rochester, Rochester, NY 14627, USA}
\author{B.Wosiek}
\affiliation{Institute of Nuclear Physics PAN, Krak\'{o}w, Poland}
\author{K.Wo\'{z}niak}
\affiliation{Institute of Nuclear Physics PAN, Krak\'{o}w, Poland}
\author{S.Wyngaardt}
\affiliation{Chemistry and C-A Departments, Brookhaven National Laboratory, Upton, NY 11973-5000, USA}
\author{B.Wys\l ouch}
\affiliation{Laboratory for Nuclear Science, Massachusetts Institute of Technology, Cambridge, MA 02139-4307, USA}

\collaboration{PHOBOS Collaboration}
% \homepage{www.phobos.bnl.gov}
\noaffiliation

\date{\today}% It is always \today, today,
\begin{abstract}
The centrality dependence of the midrapidity charged-particle
multiplicity density ($|\eta|$$<$$1$) is presented for Au+Au and Cu+Cu
collisions at RHIC over a broad range of collision energies.  The
multiplicity measured in the Cu+Cu system is found to be similar to
that measured in the Au+Au system, for an equivalent $\npart$, with
the observed factorization in energy and centrality still persistent
in the smaller Cu+Cu system.  The extent of the similarities observed
for bulk particle production is tested by a comparative analysis of
the inclusive transverse momentum distributions for Au+Au and Cu+Cu
collisions near midrapidity.  It is found that, within the 
uncertainties of the data, the ratio of yields between the various
energies for both Au+Au and Cu+Cu systems are similar and constant
with centrality, both in the bulk yields as well as a function of
\Pt, up to at least 4~GeV/$c$.  The effects of multiple nucleon
collisions that strongly increase with centrality and energy appear
to only play a minor role in bulk and intermediate transverse momentum 
particle production.
\end{abstract}

\maketitle

The experimental program at the Relativistic Heavy-Ion Collider (RHIC)
at Brookhaven National Laboratory has opened a new era in the study of
QCD matter.  Measurements of the midrapidity density of charged
particles have, from the start of the RHIC program, provided an estimate
of the energy density created in these collisions as well as an
important baseline against which to test core features of particle
production in models.  The wealth of data on bulk charged-particle
production collected in the first 5~years of RHIC operations has
revealed the presence of various features in the data that appear
to be universal, including a factorization of the charged-particle
production into two components, one that depends on collision energy,
or $\snn$, and the other on the centrality of the colliding
heavy-ions~\cite{cite:PHOBOS_White_Paper}.  The energy dependence of
the midrapidity charged-particle density appears to be consistent with
a logarithmic rise with $\snn$~\cite{cite:PHOBOS_First_200}. 
This logarithmic dependence resembles
that in elementary p+p and \ee collisions~\cite{cite:PHOBOS_pp_ee_AA}
and was not {\it a priori} expected for relativistic heavy-ion collisions.
In particular, initial expectations of a large increase in particle
production with collision energy for central heavy-ion collisions
due to multiple binary nucleon scatterings and the increased formation
of mini-jets were not realized.  Further studies revealed a relatively
weak dependence of the produced midrapidity charged-particle
multiplicity density per participant pair on the collision centrality.
From a two-component picture~\cite{cite:KharzeevNardi}, the part of
the multiplicity which scales with the number of binary collisions
($\ncoll$) contributes at the level of only $\approx$13\%, across all
energies studied so far at 
RHIC~\cite{cite:PHOBOS_200_Tracklets,cite:PHOBOS_20_Tracklets}. 

This paper reports results from RHIC for the midrapidity 
charged-particle density in Cu+Cu collisions at $\snn$ = 22.4, 62.4
and 200~GeV as a function of centrality.  The data correspond to the
top 50\% of the total inelastic cross-section, and range from an
average number of participating nucleons $\avenp$~=~20 to~108.   
The previously published study of the centrality dependence of Au+Au
multiplicity also covered the top 50\% of the total inelastic cross-section,
corresponding to $\avenp > 60$.  Thus, the full investigation of bulk
charged-particle production as a function of $\npart$ was limited by
the wide gap between the results of p+p collisions and the lowest
measured Au+Au centrality bin.  For Au+Au, this left room for
different scenarios of particle production evolution in heavy-ions,
and made it impossible to distinguish various evolution scenarios
(for example, a smooth rise or a sudden jump at low $\npart$ values).
One way to circumvent this is via a detailed mapping of the particle
production as a function of $\npart$, from values corresponding to
mid-central Au+Au collisions all the way down to those in d+Au
collisions.  The Cu+Cu data is therefore ideally suited for this
study as the average number of participants in the most central data
class ($\avenp^{\rm max} = 108$) overlaps with the peripheral Au+Au
measurements, and the average number of participants in the most
peripheral bin ($\avenp^{\rm min} = 20$) reaches down to the upper
limit for central d+Au collisions.  In addition, for the same number
of participants as in Au+Au collisions, the relative uncertainty in
the fractional cross-section of Cu+Cu collisions is smaller, enabling
a more precise measurement.  Also discussed in this paper, a subsequent
re-analysis of the Au+Au data, using the techniques developed for
Cu+Cu data analysis, extends the overlap in $\npart$ down to
$\sim$20~participants.

The new Cu+Cu collision data were recorded by the PHOBOS experiment
during the 2005 RHIC run. The primary hardware triggering was 
performed, as in prior PHOBOS heavy-ion data~\cite{cite:PHOBOS_White_Paper}, 
by two sets of 16 scintillator ``Paddle'' counter arrays that cover
a pseudorapidity of $3.2<|\eta|<4.5$ ($\eta=-\ln(\tan(\theta/2))$).
The read-out from the full PHOBOS detector was initiated by the
occurrence of one or more scintillator hits on each array within
$\Delta t<10$~ns.
This trigger prevents the loss of central collision events that may occur
for Cu+Cu collisions if the primary hardware trigger is based solely on
the zero-degree calorimeters (ZDC's), which detect spectator neutrons.
For this analysis, the energy and time information from the ZDC's were
used as an additional hardware cross-check to remove any contamination
from beam-gas background events for mid-central collisions.  The
resultant data were then passed through a series of off-line quality
checks before undertaking the physics analysis.  The two most important
of these checks include a cut on events with a time difference, between
the Paddles, of 5~ns or less to filter out beam-gas
collisions, and a further cut on events that had a valid reconstructed
primary vertex, determined as detailed below, within 10~cm of the nominal
collision position.  

The primary algorithms for vertex reconstruction in PHOBOS were
initially optimized for high multiplicity, central
Au+Au collisions.  For the Cu+Cu data, where multiplicity on average is
much lower, the nominal reconstruction procedures are found to be 
inefficient, even for mid-central data.  In order to maximize the
usability of this dataset, two new vertex reconstruction techniques
were designed to allow for the analysis of the more peripheral events.
The first method finds the vertex position from hits in the
single-layer Octagon detector ($|\eta|<3.2$) using a probabilistic
approach applied to the energy deposited at each
hit~\cite{cite:Phobos_Vertexing_NIM}.  The overall vertex resolution
attained by this technique is found to be $\sigma \approx 0.6 - 1.0$~cm,
dependent on the multiplicity of a given event. In order to obtain a
better resolution along the beam axis ($z$), a complementary method based on
all possible two-point straight tracks is used.
Straight line trajectories are assumed as the magnetic field in the region of
the vertex detector or the inner six planes of the Spectrometer (used
in the vertex reconstruction) is negligible~\cite{cite:Phobos_NIM}.
If the two-point track, formed from the Vertex detector and the first
two silicon layers of the Spectrometer, passes close to the ($x$, $y$)
orbit position of the beam, then the $z$-coordinate at the orbit position
is considered a candidate for the collision vertex. From this ensemble
of tracks, the $z$-position is determined from the most frequently found
position.  As a quality assurance measure, it is required that these two
(independent) vertex positions are within $|\Delta z|<3$~cm of each
other.  This requirement removes all mis-identified collision vertices
without detriment to the efficiency.  Agreement to within 3~cm represents
a 3$\sigma$ acceptance of the lower resolution Octagon-based vertex.  The
overall vertex resolution for this procedure is determined by the
two-point track vertex reconstruction method and is found to be
$\sigma<0.1$~cm for high multiplicity events and $\sigma\sim0.25$~cm for
low multiplicity data.   This overall improvement in vertex
reconstruction efficiency, while still maintaining a good resolution,
has allowed all track-based analyses (including those
reported here) to be carried out in the Cu+Cu collision system. 

\begin{figure*}[!t]
\centering
\includegraphics[angle=0,width=0.95\textwidth]{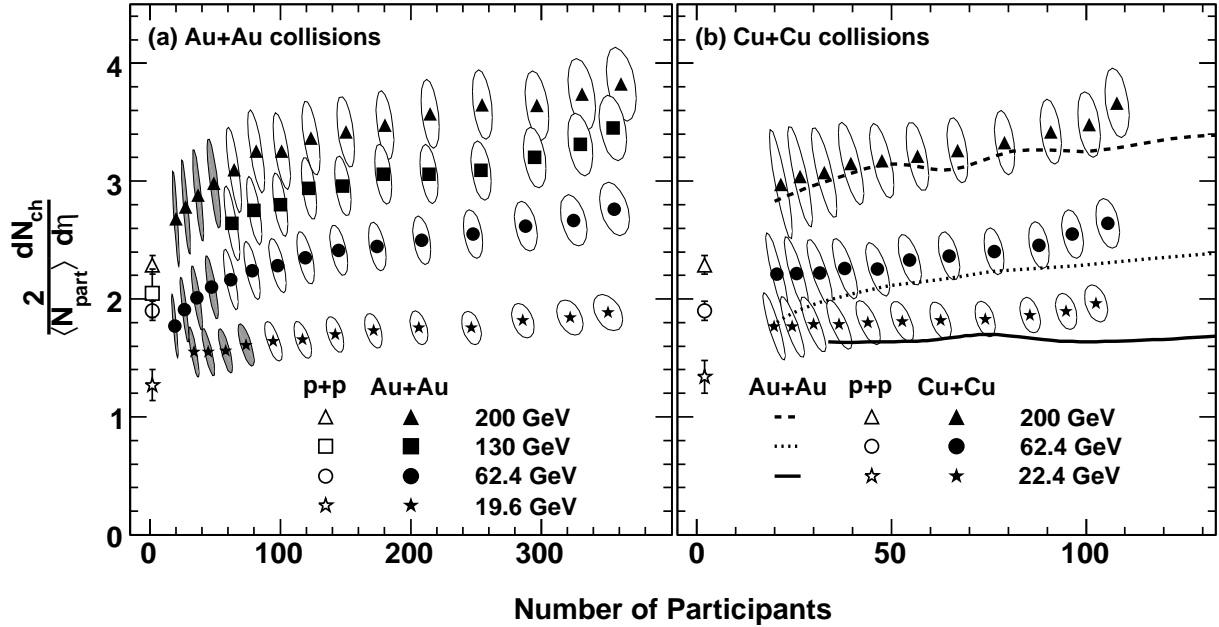}
\caption{\label{fig:dNdeta}
The charged-particle pseudorapidity density at midrapidity ($|\eta|<1$),
scaled by the number of participating nucleon pairs, for (a) Au+Au
collisions and (b) Cu+Cu collisions.  The shaded ellipses in panel (a)
represent the uncertainty on the new peripheral data, unshaded ellipses
are from Ref.~{\protect \cite{cite:PHOBOS_62_Tracklets}}. In panel (b),
the lines represent the Au+Au data (no errors) for comparison, where the
19.6~GeV data is scaled by 1.055 (derived
from~\cite{cite:PHOBOS_62_Tracklets}) to account for the difference in
collision energy of 22.4~GeV for the Cu+Cu data.  The open symbols at
$\npart=2$ represent the inelastic p+p data at equivalent energies to
the data, as in Ref.~{\protect \cite{cite:PHOBOS_20_Tracklets}}.  
The ellipses represent 90\% C.L. systematic errors.}
\end{figure*}

To provide a measure of the centrality of the collision, and subsequent
assignment of model-based parameters such as the number of participants,
the data are divided into bins of fractional cross-section.  These bins
allow for the comparison of data between collision systems and energies.
The first step toward determining the cross-section binning is to
estimate the trigger and vertex reconstruction efficiency.  Two methods
were employed for the Cu+Cu data.  The ``Paddle hit'' method counts the
number of scintillator Paddle slats hit and compares them to the number
from a fully simulated Monte Carlo (MC) sample of events.  This method
was also used in the analysis of 62.4-200~GeV Au+Au data.
The inefficiency is estimated from the loss of events in the region of
lowest number of Paddle slats hit, see for example
Ref.~\cite{cite:PHOBOS_White_Paper}.  A second technique, employed for
both 19.6~GeV Au+Au and 200~GeV d+Au collisions, known as the shape
matching method, compares the distribution of hit multiplicity in
the Octagon detector to that in the MC
simulation~\cite{cite:PHOBOS_20_Tracklets}.  The MC distribution is
scaled to match the maximum multiplicity in data (assuming 100\%
efficiency at maximum multiplicity) such that a ratio of the
multiplicity distributions in data and in MC corresponds to the
efficiency.  Both of these techniques
were used with two different MC generators
({\sc hijing}~\cite{cite:hijing} and {\sc ampt}~\cite{cite:ampt})
and {\sc geant}~\cite{cite:geant} simulations of the detector response
to determine the efficiency.  The combined trigger and vertex
reconstruction efficiencies for Cu+Cu collisions were found to be
84$\pm$5\%, 75$\pm$5\%, and 79$\pm$5\% for 200, 62.4 and 22.4~GeV
respectively.  A small difference ($\sim$5\% in determined efficiency)
was found between methods, and this was considered in the final
systematic uncertainty of the centrality determination.
For this analysis, the centrality cross-section bin selection was based
on the total energy deposited in the
Octagon detector.  As a cross-check of auto-correlation biases, the
centrality bin selection was also determined using the sum of energy in
the Paddles, truncated to the 24 (of 32) lowest energy signal slats to
remove Landau-tail fluctuations.
No difference in the results between the two centrality methods was found
for peripheral to mid-central data.  A 3\% difference was found only for
the most central data point, which is covered by the systematic
uncertainty on the measurement.  The average number of participating
nucleons, $\avenp$, for a given centrality bin was extracted from the
initial Glauber-based values in the {\sc hijing} MC simulation for the
associated centrality cuts on the final fully simulated Paddle and
Octagon centrality distributions.  Systematic uncertainties on $\avenp$ were
determined using different event generators including the uncertainties
in the efficiency of both triggering and vertex reconstruction.
For more details on the PHOBOS centrality methods see
Refs.~\cite{cite:PHOBOS_White_Paper,cite:Phobos_Centrality_Bari}.

\begin{figure*} [!t!b]%  ! means try harder
\resizebox{0.95\textwidth}{!}{\includegraphics{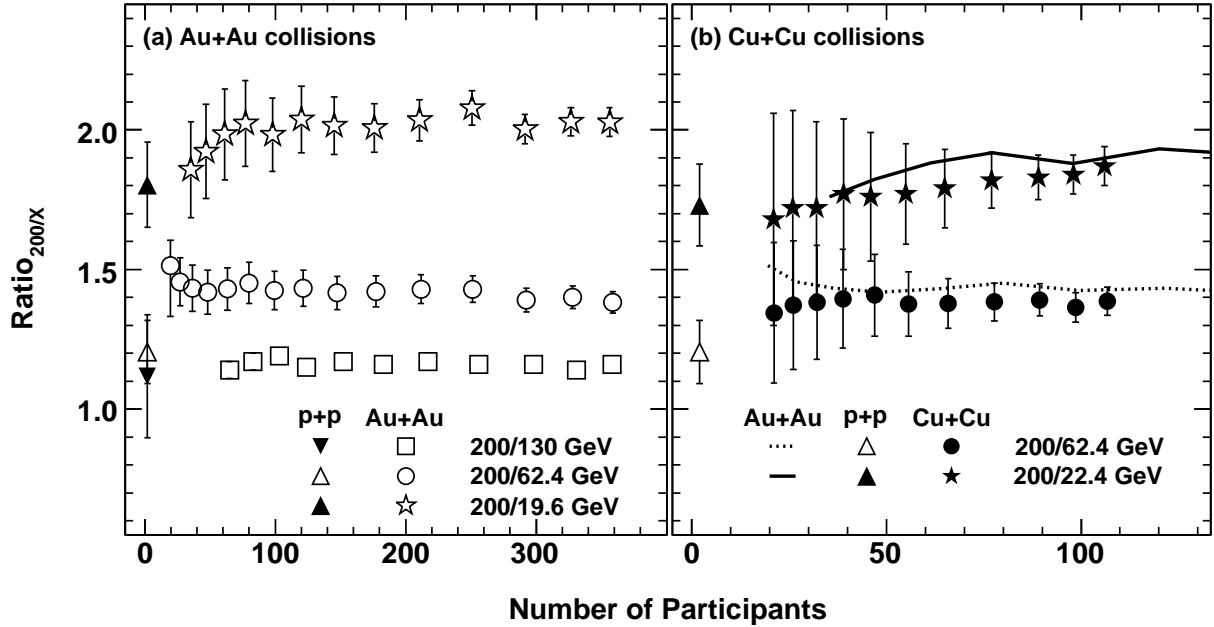}}
\caption{\label{fig:TrRatios}
The ratio between two collision energy datasets (200~GeV / $X$~GeV)
of the charged particle pseudorapidity densities from
Fig.~\ref{fig:dNdeta}.  Panel (a) shows the data ratios from
Au+Au collisions~\cite{cite:PHOBOS_62_Tracklets}.
  Panel (b), shows the Cu+Cu data, along with lines
representing the Au+Au data (no errors) for comparison.  In panel (b)
the 19.6~GeV Au+Au data is scaled by 1.055 to account for the difference
in collision energy between the low energy Au+Au (19.6~GeV) and Cu+Cu
(22.4~GeV) data.  The triangles (at $\npart=2$) represent the inelastic
p+p data at equivalent energies to the Au+Au data, as in
Ref.~{\protect \cite{cite:PHOBOS_20_Tracklets}}.  
Vertical error bars are combined statistical and systematic $1$-$\sigma$
uncertainties.}
\end{figure*}

The midrapidity charged-particle pseudorapidity density measurement
in this report is based on hit patterns in the dual-plane Vertex silicon
detector~\cite{cite:Phobos_NIM}.  Two reconstructed hits in
either the top or bottom pair of
planes, form a ``tracklet''.  Among all possible hit combinations, or
tracklets, in these two detector planes, only those that point back to
the pre-determined vertex position (within a distance of closest 
approach of 6~mm) are utilized for the final measurement, greatly reducing the
number of fake tracklet candidates.  The charged-particle multiplicity
is determined by counting the number of tracklet candidates and applying
a correction factor to account for the geometrical acceptance of the
detector, reconstruction efficiency, and vertex resolution~\cite{cite:PHOBOS_20_Tracklets}.  Possible effects on the tracklet
data, due to the changing vertex resolution with multiplicity, were
investigated with no systematic effects observed.  The results 
were further checked for possible auto-correlation biases (due to the
new vertex reconstruction procedures) via comparison to the \Pt~integrated
invariant yields~\cite{cite:PHOBOS_Spectra_Cu} and the single-layer
analysis using the Octagon detector~\cite{cite:PHOBOS_Mult_Cu}.  All
results are found to be consistent.

The charged-particle pseudorapidity density, at midrapidity, scaled by
the number of participating nucleon pairs is shown in
Fig.~\ref{fig:dNdeta} for Au+Au and Cu+Cu collisions. Panel (a) shows
the published Au+Au data~\cite{cite:PHOBOS_62_Tracklets} and for each energy
four new peripheral data points, with shaded error ellipses.  The
new Au+Au data (analyzed using the techniques developed for Cu+Cu)
extend the studies of charged particles at mid-rapidity from a fractional
cross-section of 50\% to a peripheral value of 70\%.  The triggering
and vertex reconstruction efficiency are 100\% in this region.  With
these new peripheral data it is now apparent that the charged-particle
multiplicities at midrapidity tend towards the inelastic p+p data.  In
addition, the new peripheral Au+Au data provide a large overlap in $\avenp$
with the Cu+Cu data, Fig.~\ref{fig:dNdeta}a.  The Au+Au data points are
also shown in Fig.~\ref{fig:dNdeta}b as a line, in comparison to the
Cu+Cu data.  The charged particle yields are higher in Cu+Cu collisions
than for inelastic p+p data, and, as for Au+Au collisions, the
multiplicity per participant pair increases with centrality and collision
energy.  Within the systematic uncertainty, the trends in the centrality
dependence of the charged particle production are the same across
systems.  The degree to which the trends are the same is somewhat
surprising in the context that the number of participants differ by a
factor of three for the same cross-sectional fraction, due to the much
smaller size of the Cu nucleus.

A more precise way to look at trends in the data is to take ratios
of the midrapidity charged particle yields at different collision
energies, as a function of centrality.  In this way a large fraction
of the systematic uncertainties in the measurement cancel, as
detailed in~\cite{cite:PHOBOS_20_Tracklets}, 
leaving a combined (statistical and systematic) uncertainty at the
level of a few percent.  The primary reason for this cancellation is
that the analysis uses the same detector, reconstruction technique,
triggering requirements and centrality determination method across the
different data samples.  Systematic uncertainties arising from a
common source are thus neutralized.  The result of this analysis for
all PHOBOS data on Au+Au and Cu+Cu collisions is shown in
Fig.~\ref{fig:TrRatios}.
These ratios are taken as a function of fractional cross-section and
shown versus the corresponding average $\npart$ value between the two
energies.  The small differences in $\avenp$ for the same fraction of
cross-section for different collision energies has a negligible effect 
within a given collision system.  Performing the analysis in the Au+Au
system for precisely matched $\avenp$ values yielded the same
results~\cite{cite:PHOBOS_20_Tracklets}.  Furthermore, the ratios
obtained using different centrality methods are consistent within the
shown uncertainties.

We observe two striking features in the data.  First, even within the
reduced uncertainties, the slope of the ratios are consistent with
zero for both Cu+Cu and Au+Au systems as a function of centrality.
This observation in the data clearly illustrates that the
factorization in centrality and collision energy of the 
midrapidity charged-particle yields is a feature of both the large
Au+Au system as well as the smaller Cu+Cu system.  As the ratios show no
measurable centrality dependence, due to the factorization
into energy and centrality components, the trends of $\dndetaone$
with centrality seen in Fig.~\ref{fig:dNdeta} are not driven by the
collision energy, and thus must be a consequence of the collision
geometry, i.e.~the volume of the interaction as defined by the number
of participating nucleons. 
Second, we observe that the increase in charged-particle yields with 
collision energy is similar between the Cu+Cu and Au+Au systems.  
In particular, the ratio of midrapidity charged-particle yields from
200 and 62.4~GeV are the same in Cu+Cu and Au+Au systems, within the
systematic uncertainties.  The offset in the ratio 
of 200/19.6 Au+Au compared to 200/22.4 Cu+Cu is largely 
understandable in terms of the difference in the lower collision energies
(i.e.~19.6 versus 22.4~GeV).

\begin{figure}[t]
\includegraphics[angle=0,width=0.475\textwidth]{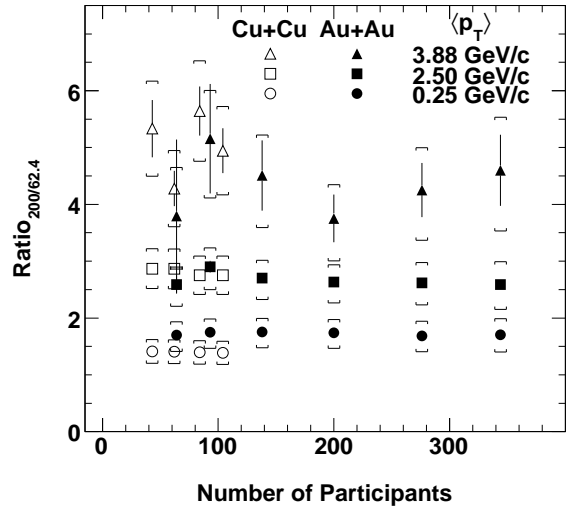}
\caption{\label{fig:PtRatios}
Ratio (200~GeV / 62.4~GeV) of charged-particle yields measured at
three different $\langle$\Pt$\rangle$~values versus centrality in Au+Au (closed symbols) and Cu+Cu (open)
collisions.  Vertical error bars represent the statistical uncertainty
in the measurement, brackets represent 90\% C.L. systematic errors.
The data are from Refs.~\cite{cite:PHOBOS_Spectra_62.4,cite:PHOBOS_Spectra_200,cite:PHOBOS_Spectra_Cu}.
}
\end{figure}

The extent of the similarities in the charged-particle ratios for
Cu+Cu and Au+Au collisions can be further tested by a comparative
analysis of the inclusive charged-particle transverse momentum (\Pt)
distributions. For this, data from Ref.~\cite{cite:PHOBOS_Spectra_62.4} 
are used to form a ratio of charged-particle yields between $\snn=200$
and 62.4~GeV as function of centrality and \Pt.  The results,
shown in Fig.~\ref{fig:PtRatios} for three bins of \Pt, are
independent of centrality, in agreement with those from the
bulk yields, even for transverse momenta up to \Pt$\sim$4~GeV/$c$. 
One may expect that the produced yields at intermediate- to
high-\Pt~would be
predominantly formed from hard partonic collisions, i.e.~those more
likely to produce jets.  As the hard collisions should follow a
scaling with $\ncoll$ then a distinct $\npart$ dependence of
the ratio should be observed for different \Pt ranges.
In this regard, it is surprising to find no
evidence of a centrality dependence apparent in the data at high-\Pt.
The increase of the average ratio value with \Pt~is expected due to
the harder spectra at 200~GeV compared to 62.4~GeV, as also observed
in p+p collisions.  A more definitive statement at high-\Pt~would
require data with higher statistics and lower systematic uncertainties.

It is important to point out that the effects observed in the
\Pt~dependence favor a volume-dominated scenario of particle production,
and it will be interesting to see whether the further differential
measurements (particle species dependent effects) support this conclusion.
Comparisons of proton to pion ratios in Au+Au collisions 
at 200 and 62.4~GeV are consistent with our observations
for unidentified charged-particles \cite{cite:STAR_200_62_pid}.
From our study, it is clear that effects of
multiple-nucleon collisions which strongly grow with centrality and energy
play only a minor role, if any, in bulk charged-particle production.

In conclusion, this paper reports new data for the midrapidity
charged-particle density in Cu+Cu and peripheral Au+Au collisions at
RHIC.  The charged particle density per participant pair is found to
be similar between peripheral Au+Au and Cu+Cu, for the same
number of participants. The midrapidity yields of the smaller Cu+Cu
system appear to factorize into collision energy and centrality
dependent components, in much
the same manner as seen in the Au+Au system.  The observed centrality
independence of the ratios, between different center-of-mass energies, further
illustrate that the collision geometry drives the bulk yield of charged
particle production at
RHIC.  This apparent general feature also appears to hold as a
function of transverse momentum for \Pt up to at least $\sim4$~GeV/$c$.

\begin{acknowledgments}

% Phobos acknowledgements and funding credits
%
% Last edited 4-Apr-2008 by George Stephans
%
% For expanded acknowledgements, you can uncomment some or all of the following
%
% We acknowledge the generous support of the Collider-Accelerator Department
% (including RHIC project personnel) and Chemistry Departments at BNL.  We
% thank Fermilab and CERN for help in silicon detector assembly.  We thank the
% MIT School of Science and LNS for financial support.  
%
This work was partially supported by U.S. DOE grants 
DE-AC02-98CH10886,
DE-FG02-93ER40802, 
DE-FG02-94ER40818,  % MIT
DE-FG02-94ER40865, 
DE-FG02-99ER41099, and
DE-AC02-06CH11357, by U.S. 
NSF grants 9603486, % Phobos TOF 
0072204,            % Rochester until 6/03
and 0245011,        % Rochester starting 6/03
by Polish MNiSW grant N N202 282234 (2008-2010),
by NSC of Taiwan Contract NSC 89-2112-M-008-024, and
by Hungarian OTKA grant (F 049823).
\end{acknowledgments}

\end{document}